\documentclass[runningheads]{llncs}

\usepackage{placeins}
\usepackage[T1]{fontenc}
\usepackage{graphicx}
\usepackage{capt-of}
\usepackage{makecell}
\usepackage{float}
\usepackage{amsmath,amssymb}
\usepackage{booktabs}
\usepackage{multirow}
\usepackage{xcolor}
\usepackage{tikz}
\usetikzlibrary{shapes.geometric, arrows.meta, positioning, fit, backgrounds, shadows}

\usepackage{tikz}
\usetikzlibrary{positioning,arrows.meta,shapes.geometric,calc,fit,backgrounds,shadows}

\colorlet{colBlue}{blue!40}
\colorlet{colProj}{teal!50}
\colorlet{colUp}{orange!50}
\colorlet{colCBAM}{purple!50}
\colorlet{colPink}{magenta!50}

\begin{document}

\title{Polyp-D2ATL: Deep Domain-Adaptive Transfer Learning for Colorectal Polyp Classification under Label Distribution Shift}
\titlerunning{Polyp-D2ATL}



\author{Sajad Jabarzadeh Ghandilu\inst{1} \and
Maryam Sadat Hosseini Azad\inst{2} \and
Shahriar Baradaran Shokouhi\inst{2} \and
Emad Fatemizadeh\inst{1}\thanks{Corresponding author}}

\authorrunning{S. J. Ghandilu et al.}

\institute{School of Electrical Engineering, Sharif University of Technology, Tehran, Iran\\
\email{fatemizadeh@sharif.edu}
\and
School of Electrical Engineering, Iran University of Science and Technology, Tehran, Iran}



\maketitle

\begin{abstract}
Early and highly accurate prediction of colorectal polyps, as an important sign of one of the most dangerous types of cancer, will result in saving more lives. Despite the advancements in colorectal polyp classification, many challenges remain in obtaining an automated polyp prediction system that is able to diagnose the difficult-to-predict polyps accompanied by different features in real scenarios, where the model can handle imbalanced data, label distribution shift, and cross-modality generalization successfully. In this study, we propose Polyp-D2ATL, a novel framework accompanied by a specific training strategy, which mitigates these limitations and effectively predicts the different classes of polyps belonging to the NICE classification. Our extensive experiments on the PICCOLO validation and test sets demonstrate that the proposed Polyp-D2ATL significantly outperforms existing state-of-the-art models across various reliable metrics, achieving an accuracy of 82.38\%, a Macro-F1 of 77.49\%, and a specificity of 87.47\% on the validation set, alongside consistent improvements on the held-out test set which demonstrates the generalization capacity and clinical applicability of the proposed approach.
\end{abstract}

\section{Introduction}

Among the various cancers worldwide, colorectal cancer is recognized as the third deadliest cancer for people \cite{ref1}. There is a possibility that a polyp may become malignant and develop into colorectal cancer over time. Therefore, early and accurate detection of polyps is crucial for the prevention of colorectal cancer. Colonoscopy is one of the most effective methods for colorectal cancer screening \cite{ref2}. NICE (Narrow-Band Imaging International Colorectal Endoscopic) classification is a standardized system for characterizing colorectal polyps. The NICE classification categorizes polyps into three types based on color, vessel pattern, and surface pattern, which helps predict histology during colonoscopy, with increasing malignancy risk from Type I to Type III. Accurate polyp classification can provide initial insights into their malignancy potential and risk level, assisting specialists in selecting the most appropriate treatment \cite{ref3}.
In the past few years, the use of artificial intelligence tools alongside colonoscopy has become a common and clinically significant process for the detection and classification of polyps, as various AI systems can reduce human error in polyp detection \cite{ref4,ref5}. Recent studies in the fields of classification and segmentation tasks have shown that deep learning methods, particularly convolutional neural networks (CNNs), have outperformed traditional image processing techniques and can be used for automated medical image analysis and disease diagnosis \cite{ref6,ref7}. CNNs learn more complex features using raw image pixels, which leads to improved accuracy and efficiency in polyp detection \cite{ref8}.
Various approaches are used for polyp classification \cite{ref9,ref10}. While existing architectures employ advanced techniques to predict different types of colorectal polyps, introducing a system that is capable of high-accuracy prediction while considering all NICE types on datasets with label distribution shift (i.e., different class distributions between train/val and test) remains unsolved. To the best of our knowledge, researchers often rely on private or local datasets (which are not severely imbalanced) to train and evaluate NICE classification \cite{ref11,ref12}, since NICE Type III typically has very limited samples, leading to drastic class imbalance and making it challenging to use public and reliable datasets such as PICCOLO \cite{ref21}. Furthermore, several works train and evaluate their models only on the two highly populated classes, which can significantly inflate ACC and Macro-F1 scores on PICCOLO \cite{ref13,ref14,ref15,ref16,ref17} or on other settings \cite{ref18}. Consequently, only a small number of studies have attempted evaluation across all three NICE categories under realistic conditions \cite{ref19,ref20}, and our model demonstrates superior performance across all metrics compared to these methods. Our main contributions are as follows:
\begin{itemize}
\renewcommand\labelitemi{\textbullet}
\item We propose Polyp-D2ATL, an attention-guided multi-scale feature fusion framework incorporating a Generalized Mean Pooling (GeM) layer for colorectal polyp classification under the NICE taxonomy, leading to robust discrimination across visually similar polyp subtypes under label distribution shift conditions.
\item We leverage a Res2Net-50 backbone pre-trained on colonoscopy-specific segmentation data from the PICCOLO and ColonGen Datasets, providing pixel-level polyp-aware feature initialization. This domain-adaptive initialization is paired with a two-phase fine-tuning strategy: full backbone freezing during warm-up, followed by progressive unfreezing with differential learning rates and Cosine Annealing scheduling, preventing catastrophic forgetting of colonoscopy-specific representations.
\item To address class imbalance, we employ Focal Loss with class-weight-based alpha and label smoothing, combined with MixUp augmentation. Furthermore, batchwise normalization is applied across both Narrow Band Imaging (NBI) and White Light Imaging (WLI) modalities as a domain-agnostic regularization strategy that reduces cross-modality intensity discrepancy without modality-specific preprocessing.
\item We evaluate Polyp-D2ATL under a rigorous protocol under various metrics to provide a more clinically meaningful assessment. Extensive experiments on both the PICCOLO validation and held-out test sets confirm the superiority of our model over existing state-of-the-art methods.
\end{itemize}

\section{Method}

\subsection{Overview of the Framework}
\begin{figure}[!t]
\centering
\includegraphics[width=\textwidth]{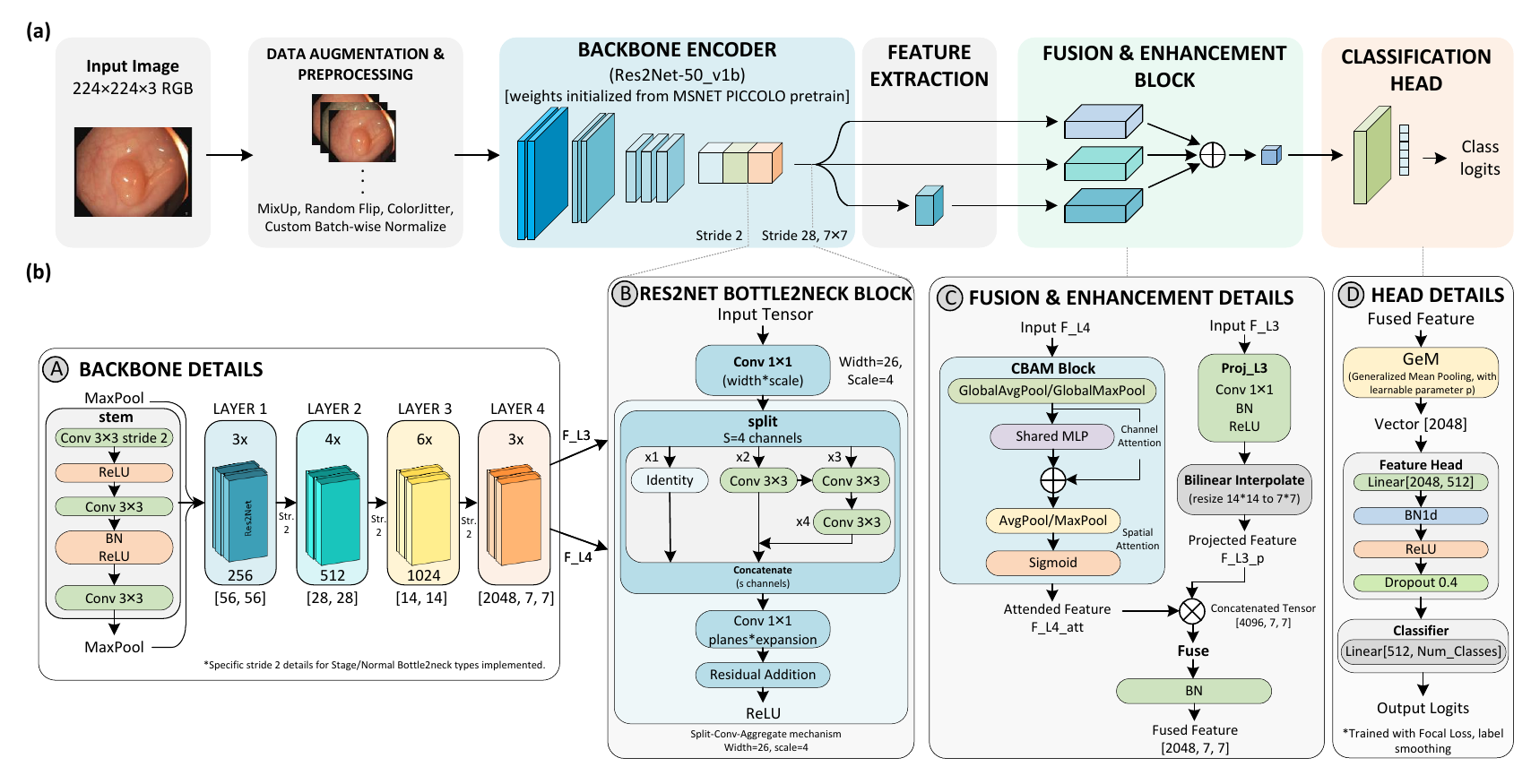}
\caption{Block diagram of the proposed domain-adaptive polyp classification framework: \textbf{(a)} the overall end-to-end flowchart of the pipeline from preprocessing to logit generation, and \textbf{(b)} comprehensive sub-component details including: (A) backbone encoder dimensions, (B) internal Res2Net Bottle2neck block design, (C) Fusion \& Enhancement details utilizing CBAM attention, and (D) deep classification head configuration.}
\label{fig:architecture}
\end{figure}

The model integrates four complementary modules: a multi-scale feature extraction backbone \cite{ref24}, a convolutional attention mechanism \cite{ref25}, a cross-layer feature fusion block \cite{ref26}, and a generalized mean pooling layer \cite{ref27}. Fig.~\ref{fig:architecture} illustrates the overall architecture of the proposed system.The proposed model follows a hierarchical feature extraction paradigm. Formally, the forward pass is defined as:
\begin{align}
&F_3, F_4 = \text{Backbone}(x), \\
&F_{\text{fused}} = \text{BN}(\text{Fusion}(\text{CBAM}(F_4), \text{Interpolate}(\text{Proj}(F_3)))), \\
&\hat{y} = \text{Classifier}(\text{FeatureHead}(\text{GeM}(F_{\text{fused}}))),
\end{align}
where $\text{Backbone}(\cdot)$ represents the feature extraction network processing the input $x$, while $F_3$ and $F_4$ are the feature maps of the third and fourth stages, respectively. $\text{Proj}(\cdot)$ denotes a $1\times1$ projection convolution, $\text{Interpolate}(\cdot)$ represents the bilinear interpolation, and $\text{CBAM}(\cdot)$ is the attention module. $\text{Fusion}(\cdot)$ signifies the concatenation-based fusion block, which is subsequently normalized by $\text{BN}(\cdot)$, the Batch Normalization operation, to generate the fused feature map $F_{\text{fused}}$. Finally, $\text{GeM}(\cdot)$ is the generalized mean pooling operation, $\text{FeatureHead}(\cdot)$ is the intermediate feature head, and $\text{Classifier}(\cdot)$ is the final classifier layer that yields the ultimate prediction $\hat{y}$.

\subsection{Res2Net-Based Multi-Scale Attention Fusion Network}
The proposed classification model is built upon a Res2Net50-v1b \cite{ref24} backbone. The backbone is initialized with domain‑specific weights derived from MSNet \cite{ref22}, which we trained on the PICCOLO dataset, transferring low-level endoscopic representations, including mucosal texture, vascular patterns, and illumination characteristics, as a stronger prior than standard ImageNet initialization. Following the final residual stage, a Convolutional Block Attention Module (CBAM) \cite{ref25} is applied to selectively refine the 2048-channel feature map in both channel and spatial dimensions. Concentrating attention exclusively at the deepest stage ensures that discriminative capacity is applied where semantic content is richest, while leaving intermediate representations intact for the subsequent fusion step. To incorporate complementary spatial detail from shallower layers, a cross-layer feature fusion block combines the attention-refined fourth-stage features $F_{\text{attention}}\in\mathbb{R}^{2048\times7\times7}$, with the third-stage features $F_3\in\mathbb{R}^{1024\times14\times14}$. The third-stage map is first projected to 2048 channels via a $1\times1$ convolution and spatially aligned to $7\times7$ through bilinear interpolation; the two tensors are then concatenated along the channel axis to form $F_{\text{cat}}\in\mathbb{R}^{4096\times7\times7}$, and a final $1\times1$ convolution followed by an additional BN layer compresses the concatenated representation back to 2048 channels, resulting in $F_{\text{fused}}$. This single-scale fusion, conceptually related to Feature Pyramid Networks \cite{lin2017feature} but simplified to a direct two-stage combination, enables the model to jointly exploit vascular and textural patterns retained in intermediate layers alongside the holistic polyp-level semantics encoded in deeper representations, a distinction particularly critical for ambiguous subtypes such as Type III polyps. The fused feature map is then aggregated using GeM. Unlike standard global average pooling, which weights all spatial locations equally, GeM adaptively shifts its behavior from spatial averaging toward spatial selection, allowing the model to emphasize diagnostically informative subregions such as pit structures and vascular patterns. The resulting 2048-dimensional descriptor is passed through a classification head consisting of a Feature Head (linear projection to 512 dimensions, BN1d, ReLU activation, and Dropout 0.4), followed by a final Classifier layer mapping to the three target classes: Type I, Type II, and Type III.

\subsection{Training Strategy and Optimization}
To prevent catastrophic forgetting of the domain-specific pre-trained representations, training proceeds in two distinct phases. In the warm-up phase (epochs 1--5), the backbone parameters are frozen, and only the classification head is optimized with a learning rate of $2\times10^{-4}$. In the fine-tuning phase, the backbone is unfrozen and trained jointly with the head for the remainder of the 100-epoch schedule. As detailed in Table~\ref{tab2}, we employ differential learning rates—$1\times10^{-5}$ for the backbone and $1\times10^{-4}$ for the head—to preserve the learned endoscopic features while adapting to the classification task.

The class distribution in the training set is imbalanced: Type II samples constitute the majority, while Type III samples represent the minority class. This imbalance is explicitly addressed through the Focal Loss \cite{ref28}, with per-class inverse-frequency weights computed as:
\vspace{0.3cm}
\begin{equation}
\omega_c=\left(\frac{1/n_c}{\sum_{i=1}^{C}(1/n_i)}\right)\cdot C,
\vspace{0.3cm}
\end{equation}
where $C=3$ is the number of NICE classes, and $n_c$ is the count of samples in class $c$. These weights are incorporated as the $\alpha$ parameters of the loss function. The total loss for the network is defined as:
\vspace{0.3cm}
\begin{equation}
\mathcal{L}_{\text{FL}}=-\alpha_t(1-p_t)^{\gamma}\ \mathcal{L}^{\text{LS}}_{\text{CE}}(p_t),
\vspace{0.3cm}
\end{equation}
where $p_t$ is the model's estimated probability for the ground-truth class, $\gamma=2.0$ is the focusing parameter, and LS denotes label smoothing with $\epsilon=0.05$ to improve calibration and prevent overconfident predictions. Optimization is performed using the AdamW optimizer with a weight decay of $1\times10^{-4}$, and the learning rate is decayed using a Cosine Annealing scheduler with $T_{\max}=100$.

Training images undergo standard geometric augmentations, including random horizontal and vertical flipping, random rotation up to $10^{\circ}$, and color jitter, all resized to $224\times224$. To further regularize the model, MixUp augmentation \cite{ref29} is applied with a probability of 0.6 and a mixing coefficient $\lambda \sim \text{Beta}(0.4, 0.4)$, training the model on convex combinations of image pairs.

Rather than using fixed ImageNet statistics, batch-wise normalization \cite{ref30} is applied at each step, computing the mean and standard deviation per channel from each mini-batch on-the-fly. This ensures that the normalized input distribution reflects the actual photometric characteristics of endoscopic images, which differ substantially from natural images in terms of color temperature and the distinct tonal properties of NBI illumination. As shown in Table~\ref{tab2}, both training and evaluation phases maintain a consistent batch size of 64 to ensure stable normalization behavior.

\FloatBarrier

\noindent\textbf{Dataset.} To evaluate the performance of Polyp-D2ATL under clinically realistic scenarios, two benchmark datasets are utilized. The primary focus of this study is the PICCOLO dataset \cite{ref21}, a comprehensive multi-modality collection comprising 3,433 manually annotated images (2,131 WLI and 1,302 NBI) derived from 76 lesions across 40 patients. This dataset serves as the core benchmark for evaluating the NICE taxonomy classification. The detailed class distribution across the train, validation, and test splits is summarized in Table~\ref{tab1}.

\begin{figure}[H]
    \centering
    \includegraphics[width=0.8\textwidth]{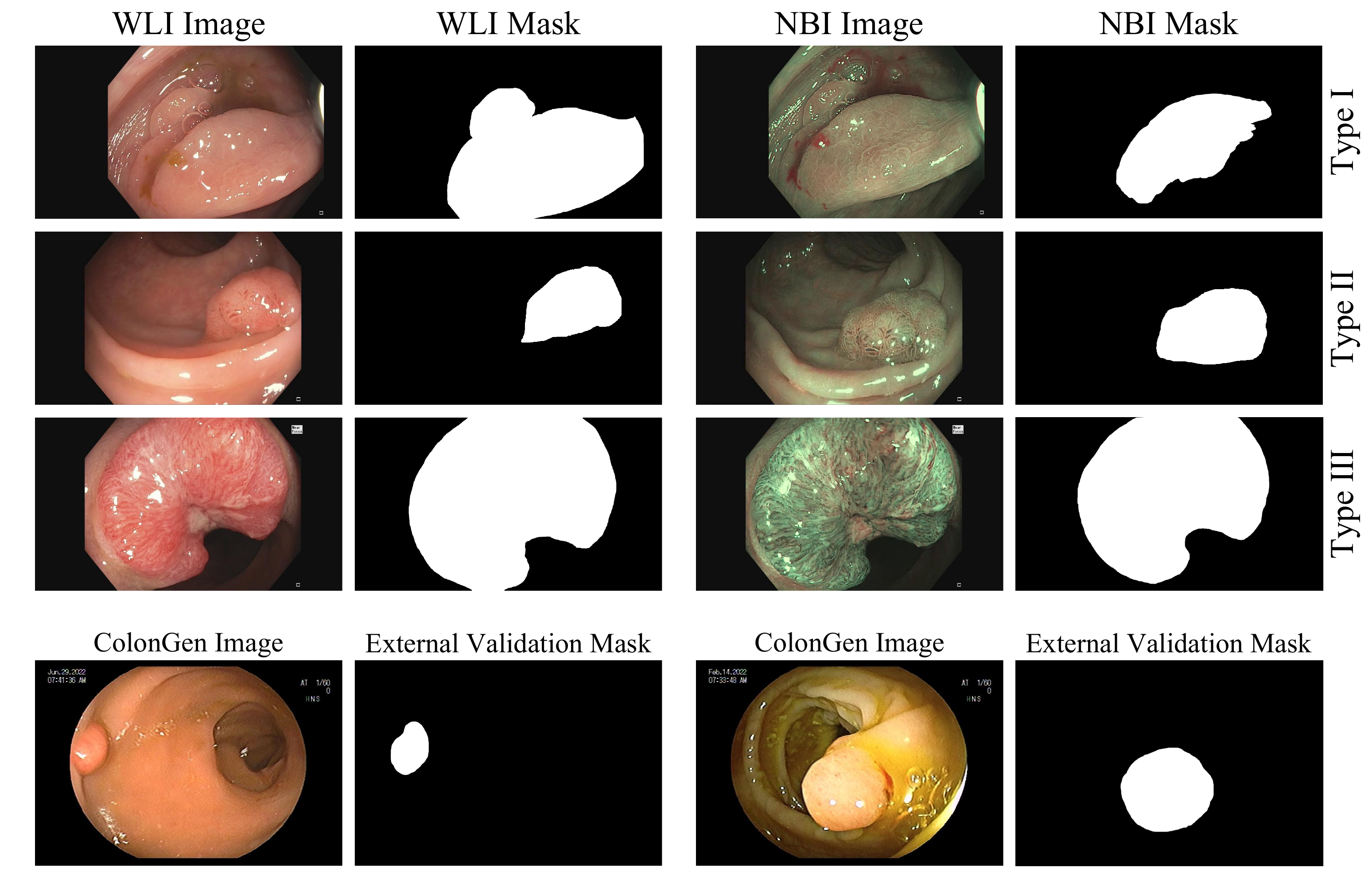}
    \caption{Visual overview of the utilized datasets. The first three rows present paired WLI and NBI samples across different NICE categories from the PICCOLO dataset. The last row illustrates independent validation samples from the ColonGen dataset, utilized to verify the generalizability of the backbone.}
    \label{fig:dataset_samples}
\end{figure}

To further verify the structural robustness and generalizability of the feature extraction backbone, the ColonGen dataset \cite{ref23} is employed as an independent external validation benchmark. ColonGen is a comprehensive polyp segmentation collection compiled from eight publicly available datasets across seven countries and twelve medical centers, specifically designed to address the geographic and equipment-based limitations of single-center studies. The high performance achieved on this heterogeneous data confirms that the proposed architecture captures universal pathological cues, ensuring that the pre-trained weights provide a reliable initialization for the subsequent classification task on PICCOLO. The visual characteristics for both datasets are illustrated in Fig.~\ref{fig:dataset_samples}.

\begin{table}[H]
\caption{Class distribution of NICE categories in the PICCOLO dataset across training, validation, and test sets.}
\label{tab1}
\centering
\renewcommand{\arraystretch}{1.0} 
\setlength{\tabcolsep}{12pt}
\begin{tabular}{|c|c|c|c|}
\hline
\multirow{2}{*}{\textbf{NICE Categories}} & \multicolumn{3}{c|}{\textbf{Number of samples in class}} \\
\cline{2-4}
 & \textbf{Train} & \textbf{Validation} & \textbf{Test} \\
\hline
Type I   & 435  & 139 & 114 \\
\hline
Type II  & 1552 & 592 & 92  \\
\hline
Type III & 172  & 166 & 127 \\
\hline
\end{tabular}
\end{table}

\noindent\textbf{Visual Variability.} As shown in Fig.~\ref{fig:dataset_samples}, PICCOLO provides modality-specific annotations to capture subtle texture changes. Furthermore, the successful representation of ColonGen samples demonstrates that our pre-trained weights are robust and generalize across diverse imaging hardware and geographic locations.

\noindent\textbf{Evaluation Metrics.} Model performance is evaluated using a comprehensive set of metrics suited to the multi-class imbalanced classification and segmentation setting. For classification, all metrics are computed in macro-averaged form to give equal weight to each class regardless of its frequency, ensuring that minority class performance is not masked by majority class results, namely, ACC, Macro-averaged Precision, Recall, and F1-score, Matthews Correlation Coefficient (MCC), and Specificity. For segmentation evaluation, mIoU, mDice, and $S_{\alpha}$ \cite{ref31} are employed. These metrics provide comprehensive insights into the effectiveness and reliability of the model.

\subsection{Implementation Details}
The configuration of the environment details is mentioned in Table~\ref{tab2}. Note that all random seeds for Python, NumPy, and PyTorch are fixed prior to training to ensure deterministic data loading and weight initialization, subject to the non-determinism inherent in CUDA parallel operations.

\begin{table}[H]
\caption{Experimental environment and hyper-parameter configuration for Polyp-D2ATL.}
\label{tab2}
\centering
\renewcommand{\arraystretch}{0.95} 
\setlength{\tabcolsep}{6pt}
\begin{tabular}{|l|l|}
\hline
\textbf{Component} & \textbf{Specification} \\
\hline
Deep Learning Framework & PyTorch 2.x with CUDA \\
\hline
Mixed Precision & torch.amp.autocast + GradScaler (float16) \\
\hline
Input Resolution & $224 \times 224 \times 3$ \\
\hline
Normalization & Batch-wise (domain-specific statistics) \\
\hline
Pre-training & Domain-adaptive MSNet (PICCOLO \& ColonGen) \\
\hline
Batch Size & 64 (consistent across all phases) \\
\hline
Optimizer & AdamW (weight decay = $1\times10^{-4}$) \\
\hline
LR Scheduler & CosineAnnealingLR ($T_{\max}=100$) \\
\hline
Warm-up Phase & 5 epochs (Backbone frozen, Head LR = $2\times10^{-4}$) \\
\hline
Fine-tuning LR & Backbone: $1\times10^{-5}$ / Head: $1\times10^{-4}$ \\
\hline
Regularization & MixUp ($p=0.6, \lambda=0.4$), Label Smoothing ($\epsilon=0.05$) \\
\hline
Total Epochs & 100 \\
\hline
DataLoader Workers & 8 (pin\_memory=True, persistent\_workers=True) \\
\hline
\end{tabular}
\end{table}

\FloatBarrier

\section{Results}

\subsection{Segmentation Pre-training}
\textbf{Segmentation.} To leverage domain-specific visual representations, we first pre-trained an MSNet backbone on the PICCOLO and ColonGen Comprehensive datasets for polyp segmentation. As reported in Table~\ref{tab3}, the backbone acquired robust colonoscopy-specific feature representations suitable for downstream classification. The quantitative performance on PICCOLO ($\text{mDice} = 0.792$) surpasses the multi-center ColonGen benchmark, primarily due to the modality-specific consistency (WLI/NBI) in the primary dataset, which facilitates sharper boundary delineation.

\begin{table}[H]
\caption{Quantitative results of MSNet segmentation on ColonGen and PICCOLO datasets.}
\label{tab3}
\centering
\renewcommand{\arraystretch}{1.2}
\setlength{\tabcolsep}{8pt}
\begin{tabular}{|c|c|c|c|c|c|}
\hline
\multicolumn{3}{|c|}{\textbf{ColonGen Comprehensive Dataset}} & \multicolumn{3}{c|}{\textbf{PICCOLO Dataset}} \\
\hline
\textbf{mDice} & \textbf{mIoU} & $\mathbf{S_{\alpha}}$ & \textbf{mDice} & \textbf{mIoU} & $\mathbf{S_{\alpha}}$ \\
\hline
0.732 & 0.653 & 0.818 & 0.792 & 0.745 & 0.859 \\
\hline
\end{tabular}
\end{table}

\vspace{15pt}

\begin{figure}[p]
    \centering
    \includegraphics[width=\textwidth]{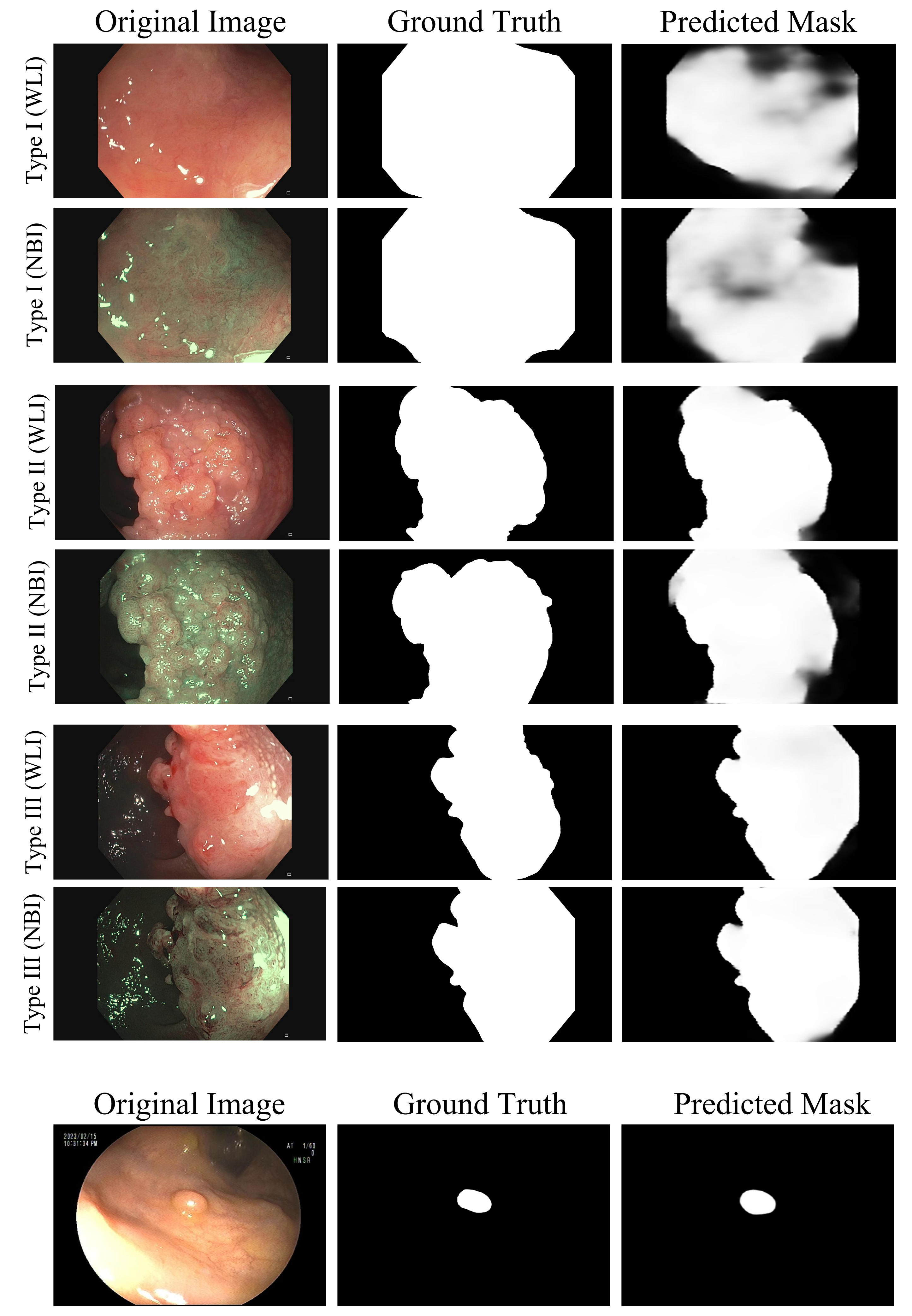}
    \caption{Qualitative segmentation performance of the pre-trained MSNet backbone. The first six rows display original images, ground truth annotations, and predicted masks from the PICCOLO dataset across different NICE categories (Type I-III) and imaging modalities (WLI/NBI). The last row illustrates an independent validation sample from the multi-center ColonGen dataset. Overall, the predicted masks demonstrate high topological fidelity, effectively delineating polyp boundaries.}
    \label{fig:qualitative_segmentation}
\end{figure}

\vspace{10pt}

\noindent\textbf{Visual Analysis.} The qualitative results in Fig.~\ref{fig:qualitative_segmentation} validate the robustness of the purposed model. In particular, the segmentation of NICE Type III polyps, which exhibit highly irregular surface textures, remains consistent across modalities. The high overlap between the predicted masks and ground truths confirms that the learned feature representations are domain-agnostic.

\subsection{Classification Performance}

\textbf{Validation Set.} As shown in Table~\ref{tab4}, Younas et al. \cite{ref19} proposed a weighted average ensemble framework combining multiple pre-trained CNN architectures with optimized hyperparameter settings for colorectal polyp classification on the PICCOLO dataset. In contrast, the superiority of our model over the previous one in all metrics shows strong discriminative capability and notably high specificity, which is critical in clinical screening scenarios to minimize false-positive detections.

\nopagebreak[4]
\begin{table}[H]
\caption{Classification performance comparison on the PICCOLO validation set. \textbf{Bold} values indicate the best results per metric.}
\label{tab4}
\centering
\renewcommand{\arraystretch}{1.15}
\setlength{\tabcolsep}{6pt}
\resizebox{\columnwidth}{!}{%
\begin{tabular}{|l|c|c|c|c|c|c|}
\hline
\textbf{Networks} & \textbf{Acc} & \textbf{MCC} & \textbf{Macro-F1 Score} & \textbf{precision} & \textbf{recall} & \textbf{specificity} \\
\hline
Younas et al. \cite{ref19} & 0.744 & -- & 0.742 & 0.782 & 0.713 & -- \\
\hline
Polyp-D2ATL & \textbf{0.823} & \textbf{0.648} & \textbf{0.774} & \textbf{0.796} & \textbf{0.766} & \textbf{0.874} \\
\hline
\end{tabular}%
}
\end{table}

\noindent\textbf{Test Set.} On the held-out test set, El-Shimy et al. \cite{ref20} proposed a modified CapsNet architecture \cite{sabour2017dynamic} pre-trained via self-supervised learning on the PICCOLO dataset. According to Table 5, the marked gap between the baseline and our method on the test set underscores the benefit of domain-adaptive pre-training combined with attention-guided multi-scale feature aggregation.

\nopagebreak[4]
\begin{table}[H]
\caption{Classification performance comparison on the PICCOLO test set. \textbf{Bold} values indicate the best results per metric.}
\label{tab5}
\centering
\renewcommand{\arraystretch}{1.15}
\setlength{\tabcolsep}{6pt}
\resizebox{\columnwidth}{!}{%
\begin{tabular}{|l|c|c|c|c|c|c|}
\hline
\textbf{Networks} & \textbf{Acc} & \textbf{Macro-F1 Score} & \textbf{MCC} & \textbf{precision} & \textbf{recall} & \textbf{SPECIFICITY} \\
\hline
El-Shimy et al. \cite{ref20} & 0.40 & 0.34 & 0.22 & -- & -- & 0.70 \\
\hline
Polyp-D2ATL & \textbf{0.54} & \textbf{0.53} & \textbf{0.45} & \textbf{0.76} & \textbf{0.57} & \textbf{0.79} \\
\hline
\end{tabular}%
}
\end{table}

\noindent\textbf{Confusion Matrix.} Fig.~\ref{fig2} shows the confusion matrix obtained by the proposed model on the PICCOLO validation set. On the validation set, the model correctly classified 519 out of 592 samples of class Type II, demonstrating strong performance on the dominant class.

\begin{center}
\includegraphics[width=0.60\textwidth]{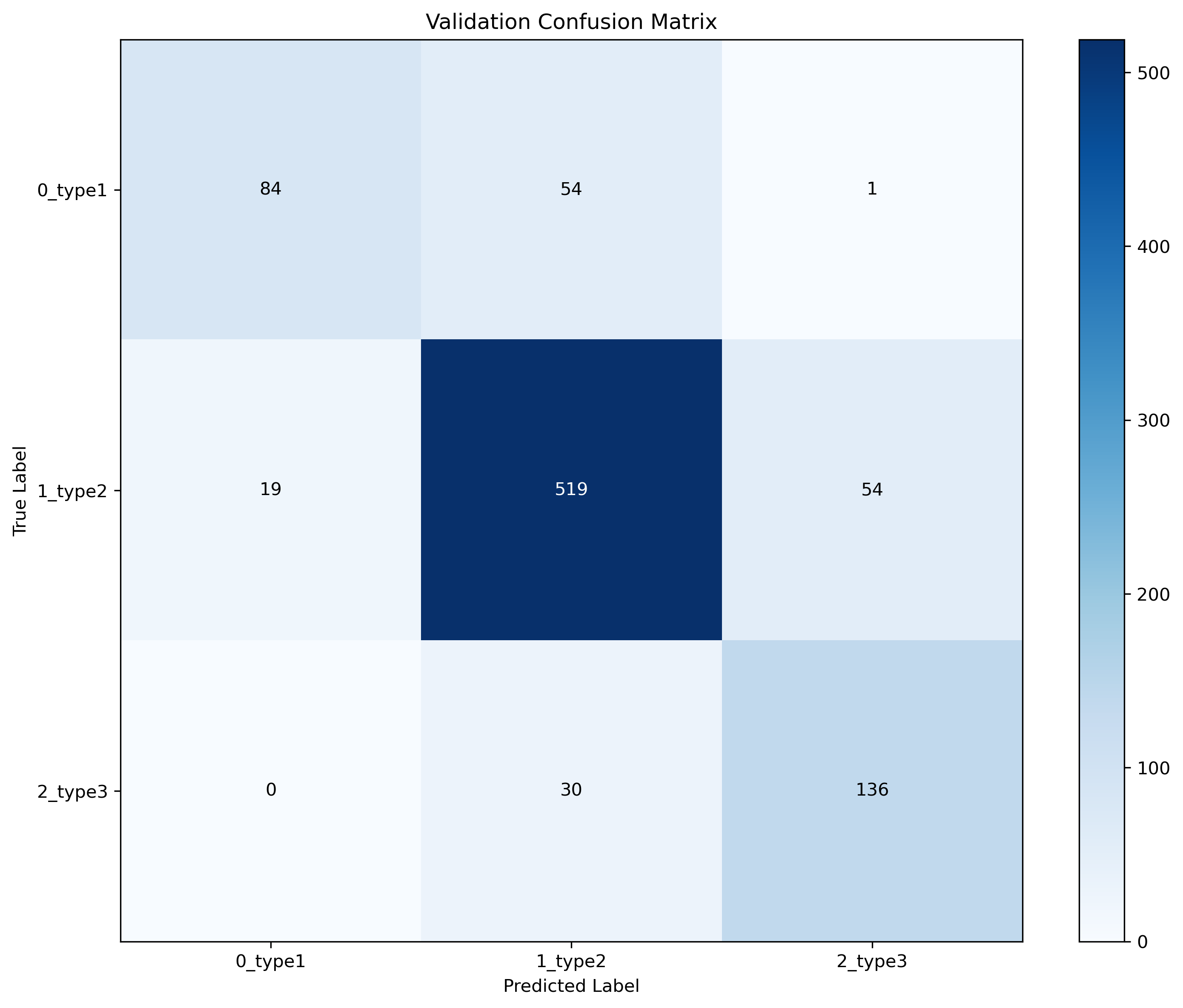}
\captionof{figure}{Confusion matrix of the proposed model on the PICCOLO validation set across three polyp classes: Type I, Type II, and Type III.}
\label{fig2}
\end{center}

\FloatBarrier

\subsection{Ablation Study}
To systematically evaluate the contribution of each proposed component, we conducted a series of incremental ablation experiments on the PICCOLO validation set. Starting from a baseline Res2Net-50 with a standard MLP classification head, we progressively added Batchwise Normalization with MixUp augmentation, GeM pooling with CBAM attention, multi-scale feature fusion, and finally Focal Loss. The results are summarized in Table~\ref{tab6}.

\begin{table}[htbp]
\caption{Ablation study results on the PICCOLO validation set, demonstrating the incremental contribution of each proposed component. \textbf{Bold} values indicate the best results per metric.}
\label{tab6}
\centering
\renewcommand{\arraystretch}{1.2}
\setlength{\tabcolsep}{6pt}
\resizebox{\textwidth}{!}{%
\begin{tabular}{|c|c|c|c|c|c|c|c|c|c|c|}
\hline
\makecell{\textbf{Res2net}\\\textbf{+}\\\textbf{MLP}\\\textbf{(baseline)}} &
\makecell{\textbf{Batchwise}\\\textbf{Norm.}\\\textbf{+}\\\textbf{MixUp}} &
\makecell{\textbf{GeM}\\\textbf{+}\\\textbf{CBAM}} &
\makecell{\textbf{Feature}\\\textbf{Fusion}} & 
\makecell{\textbf{Focal}\\\textbf{Loss}} &
\textbf{Acc} & \makecell{\textbf{Macro-}\\\textbf{F1}\\\textbf{Score}} &
\textbf{Precision} & \textbf{Recall} & \textbf{MCC} & \textbf{Specificity} \\
\hline
$\ast$ &  &  &  &  & 0.762 & 0.637 & 0.699 & 0.607 & 0.487 & 0.802 \\
\hline
$\ast$ & $\ast$ &  &  &  & 0.779 & 0.642 & 0.740 & 0.606 & 0.522 & 0.805 \\
\hline
$\ast$ & $\ast$ & $\ast$ &  &  & 0.793 & 0.637 & \textbf{0.811} & 0.615 & 0.561 & 0.812 \\
\hline
$\ast$ & $\ast$ & $\ast$ & $\ast$ &  & 0.765 & 0.676 & 0.704 & 0.657 & 0.513 & 0.821 \\
\hline
$\ast$ & $\ast$ & $\ast$ & $\ast$ & $\ast$ & \textbf{0.823} & \textbf{0.774} & 0.796 & \textbf{0.766} & \textbf{0.648} & \textbf{0.874} \\
\hline
\end{tabular}%
}
\end{table}

\noindent To further justify our choice of Focal Loss, we compared three strategies for addressing class imbalance under the full model architecture: Focal Loss (proposed), random weight sampler, and balanced batch sampling.

\begin{table}[htbp]
\caption{Comparison of class-imbalance handling strategies under the full model architecture on the PICCOLO validation set. \textbf{Bold} values indicate the best results per metric.}
\label{tab7}
\centering
\renewcommand{\arraystretch}{1.2}
\setlength{\tabcolsep}{10pt}
\resizebox{\textwidth}{!}{%
\begin{tabular}{|l|c|c|c|c|c|c|c|}
\hline
\textbf{Strategy} & \textbf{Loss} & \textbf{Acc} & \textbf{Macro-F1 Score} & \textbf{Precision} & \textbf{Recall} & \textbf{MCC} & \textbf{Specification} \\
\hline
Focal Loss (proposed) & \textbf{0.134} & 0.823 & \textbf{0.774} & 0.796 & 0.766 & 0.648 & 0.874 \\
\hline
Random weight sampler & 0.199 & \textbf{0.832} & 0.765 & \textbf{0.807} & 0.735 & \textbf{0.653} & 0.867 \\
\hline
Balanced batch sampling & 0.867 & 0.807 & 0.748 & 0.736 & \textbf{0.769} & 0.637 & \textbf{0.884} \\
\hline
\end{tabular}%
}
\end{table}

\noindent Table \ref{tab7} presents the performance of several metrics under these strategies. Focal loss achieves the lowest validation loss and the highest Macro-F1 score; meanwhile, it demonstrates comparable performance across other metrics. 

\begin{figure}[htbp]
    \centering
    \includegraphics[width=0.85\textwidth]{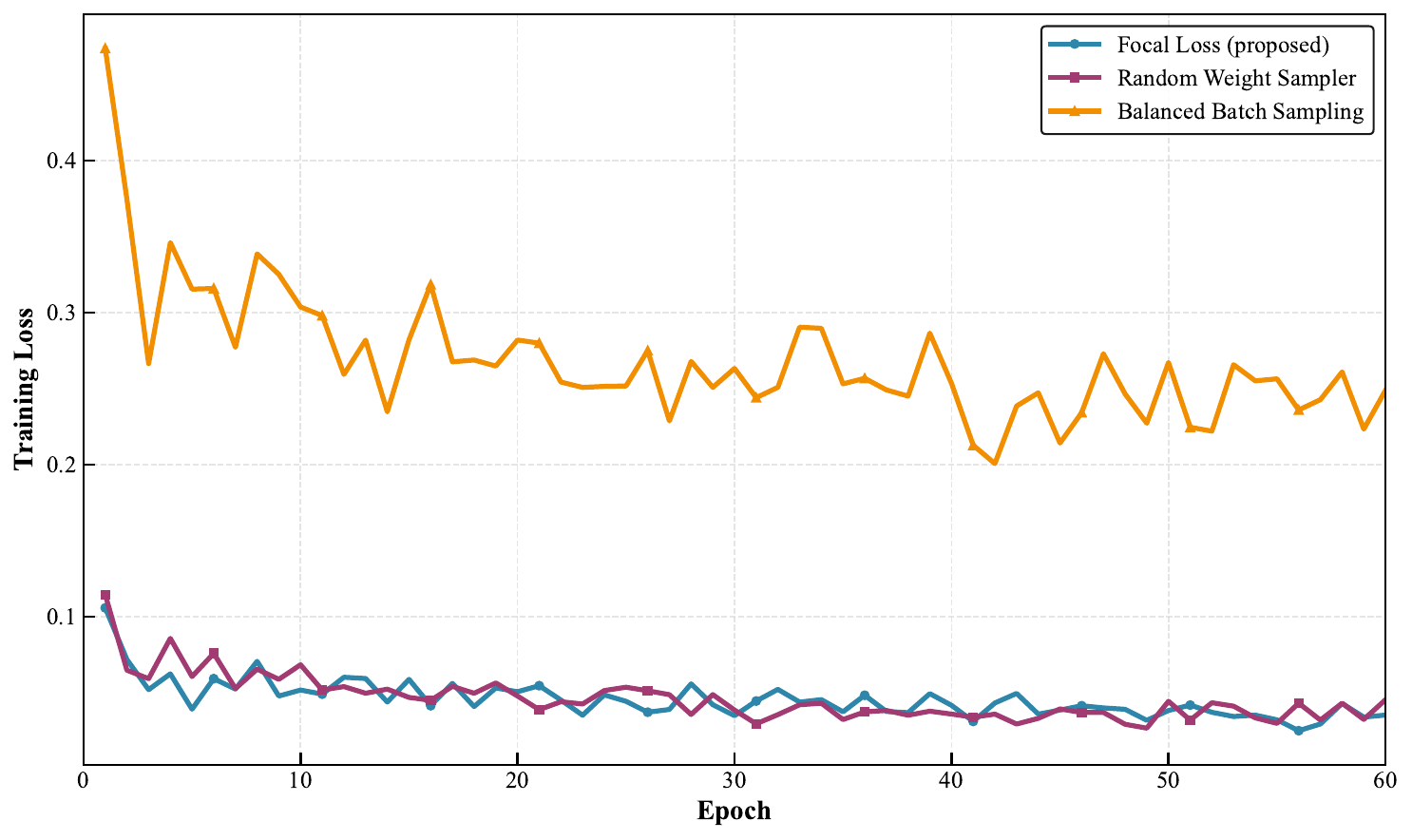}\\[1ex]
    \small (a) Training Dynamics\\[3ex]
    
    \includegraphics[width=0.85\textwidth]{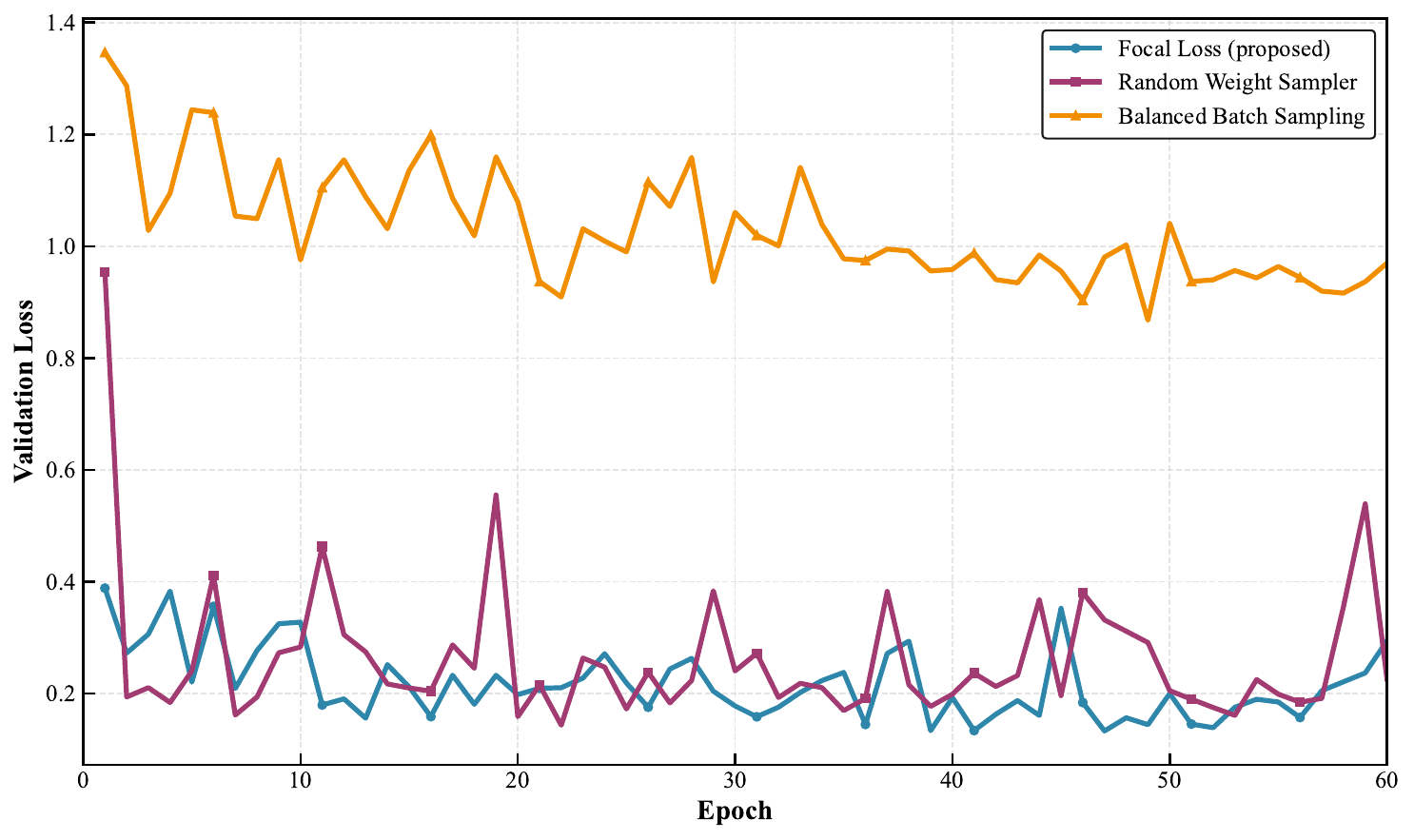}\\[1ex]
    \small (b) Validation \& Generalization
    
    \caption{Loss comparison across training epochs. (a) Training Loss shows convergence behavior and optimization stability. (b) Validation Loss highlights generalization capability and robustness against label distribution shifts.}
    \label{fig:loss_combined}
\end{figure}

Furthermore, to explicitly evaluate the generalization capacity and over-fitting susceptibility of these methods, Fig. \ref{fig:loss_combined} illustrates both the training and validation loss trajectories. As depicted in Fig. \ref{fig:loss_combined}(a), both the Focal Loss and the Random Weight Sampler achieve low and stable training losses. However, the validation trajectory in Fig. \ref{fig:loss_combined}(b) reveals that the Random Weight Sampler exhibits significant instability and sharp convergence spikes, indicating a tendency to overfit on the training set and a lack of robustness on unseen data. Conversely, Balanced Batch Sampling suffers from a consistently high loss across both phases. In contrast, the proposed Focal Loss not only successfully minimizes the training error but also maintains superior validation stability, validating its effectiveness in handling severe label distribution shifts without compromising generalization.

\section{Conclusion}
In this work, we presented a domain-adaptive deep learning framework for polyp classification, designed to address the core challenges of colonoscopy image analysis: class imbalance, label distribution shift, and cross-modality generalization. Experimental results on the PICCOLO validation and Test set demonstrated that our full model surpass the reference method in all metrics,  confirming the generalization capability of the proposed framework. Ablation experiments validated the incremental contribution of each component, with Focal Loss proving to be the most effective strategy for handling class imbalance in terms of overall F1 and training stability. These findings suggest that combining colonoscopy-specific pre-training with attention-guided multi-scale feature aggregation and robust loss design constitutes a promising direction for clinically deployable polyp characterization systems.

\subsection*{Acknowledgement}
The authors would like to thank the Basque Biobank (Biobanco Vasco) for providing access to the PICCOLO dataset.

\subsection*{Disclosure of Interests}
The authors have no competing interests to declare that are relevant to the content of this article.

\end{document}